# Enhancing frozen histological section images using permanent-section-guided deep learning with nuclei attention

Elad Yoshai[1], Gil Goldinger[2], Miki Haifler[2,3] and Natan T. Shaked[4],*

[1] School of Electrical and Computer Engineering, Tel Aviv University, Tel Aviv, Israel
[2] Chaim Sheba Medical Center, Ramat Gan, Israel
[3] Sackler Faculty of Medicine, Tel Aviv University, Tel Aviv, Israel
[4] Department of Biomedical Engineering, Tel Aviv University, Tel Aviv, Israel
* Corresponding author: nshaked@tau.ac.il

## Abstract

**In histological pathology, frozen sections are often used for rapid diagnosis during surgeries, as they can be produced within minutes. However, they suffer from artifacts and often lack crucial diagnostic details, particularly within the cell nuclei region. Permanent sections, on the other hand, contain more diagnostic detail but require a time-intensive preparation process. Here, we present a generative deep learning approach to enhance frozen section images by leveraging guidance from permanent sections. Our method places a strong emphasis on the nuclei region, which contains critical information in both frozen and permanent sections. Importantly, our approach avoids generating artificial data in blank regions, ensuring that the network only enhances existing features without introducing potentially unreliable information. We achieve this through a segmented attention network, incorporating nuclei-segmented images during training and adding an additional loss function to refine the nuclei details in the generated permanent images. We validated our method across various tissues, including kidney, breast, and colon. This approach significantly improves histological efficiency and diagnostic accuracy, enhancing frozen section images within seconds, and seamlessly integrating into existing laboratory workflows.**

## Introduction

In histopathology, frozen sections are commonly used to obtain immediate diagnosis during surgeries or other procedures, as they are fast and inexpensive to process. However, frozen sections suffer from poor image quality and artifacts and might not provide enough information for a definitive diagnosis, as the tissue may not be fully preserved, and certain features might be difficult to identify [1-4]. On the other hand, permanent sections are typically of higher quality than frozen sections, with fewer artifacts and better preservation of tissue architecture. This can provide a more detailed analysis of the tissue, allowing for a more accurate diagnosis and potential identification of additional features. However, permanent sections require much more preparation time and typically cannot be processed during surgery, which may delay diagnosis and treatment. Therefore, developing a method to improve frozen section images is critical. The preparation of frozen sections involves rapidly freezing the tissue sample to preserve its structure and cellular details. This is typically done using a cryostat, which maintains the tissue at temperatures around –20 to –30°C. The tissue is embedded in a gel-like medium, which consists of polyethylene glycol and polyvinyl alcohol. This medium ensures that the tissue remains stable and can be sliced into thin sections using a microtome. The rapid freezing process helps prevent the formation of most ice crystals that could damage the tissue, but it may still result in some artifacts and lower image quality compared to permanent sections.

In contrast, the preparation of permanent sections involves a more elaborate chemical process. The tissue is first fixed in formalin to preserve its structure and prevent degradation. It is then dehydrated through a series of alcohol baths, cleared with a solvent-like xylene, and finally embedded in paraffin wax. Once the wax hardens, the tissue block is sliced into thin sections using a microtome. These sections are then mounted on slides and stained to highlight different cellular components. The formalin fixation and paraffin embedding process provide excellent preservation of tissue architecture and cellular details, resulting in high-quality images suitable for detailed analysis.

Since frozen and permeant section chemical preparation processes are different, most available datasets include unpaired frozen and permanent section images, which makes the problem of learning and mapping between frozen to permanent section images challenging as there is no ground truth of a paired permanent section to compare to frozen sections often contain blank regions, it is crucial not to introduce artificial data into these areas, as this could mislead pathologists. An example of frozen and permanent section images can be found in Figure 1. As can be seen, the sub-nuclei features are lost in the frozen section image in comparison to the permanent section image. The nuclei texture is useful for cancer diagnosis, grading as well in prognosis prediction [5]. For example, the detailed analysis of nuclear morphometry and chromatin texture has been shown to provide critical insights into the diagnosis and prognosis of hepatocellular carcinoma [6], highlighting the importance of precise nuclear characteristics in enhancing the accuracy of pathological assessments. These analyses can be done with permanent sections and cannot be done with frozen sections.

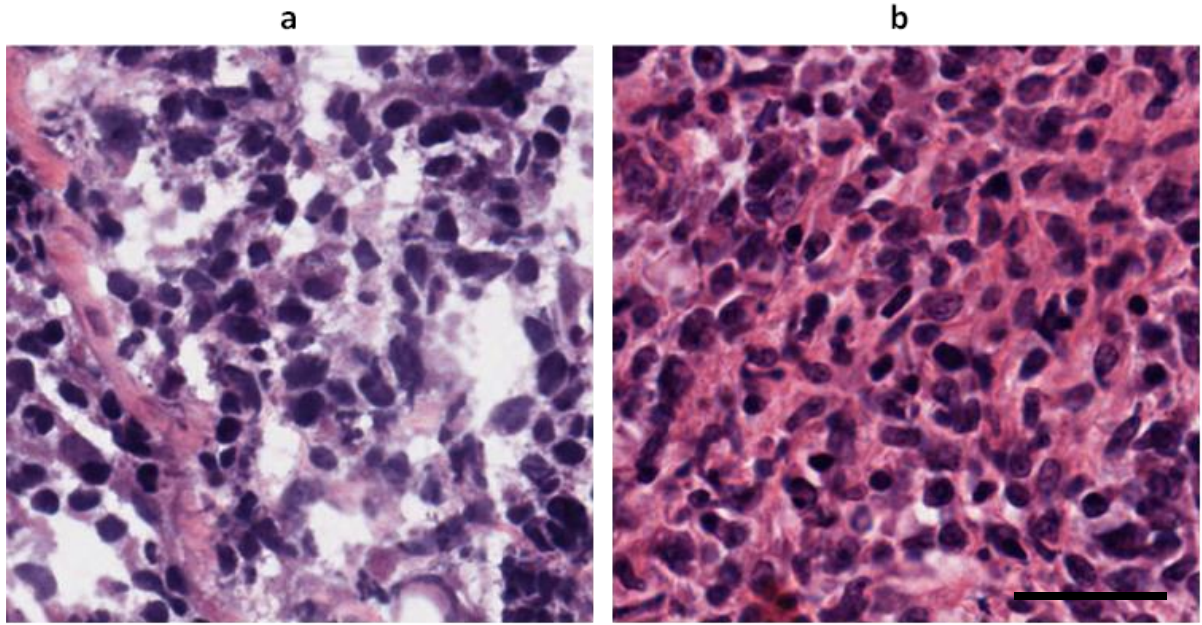

**Figure 1.** Examples of frozen (a) and permanent (b) section images. The frozen section has blank parts in the image, and the nuclei region almost does not contain details, compared to the permanent section. Scale bar: 50 µm.

With the rise of the generative adversarial networks (GANs) [7], several methods have been proposed for image-to-image translation, such as Pix2Pix [8] for paired image-to-image translation and StarGAN [9] for multi-domain image-to-image translation. To address the unpaired image-to-image translation other methods, such as the CycleGAN [10] and UNIT [11], were proposed and have shown a great promise in transferring images from one domain to another, also in the medical domain [12-16]. Recently, diffusion models [17] were adapted for the task of image-to-image translation [18-20].

In the histopathology domain, a few works have generated permanent section from frozen section images [21] using CycleGAN with perceptual loss [22], contrastive unpaired translation (CUT) [23], and stain normalization (SN) [24]. Specifically, Ozyoruk et al. [25] suggested a GAN-based model including a spatial attention block (SAB) in the generator and self-regularization (SR) loss to preserve clinically relevant features and retain the content of the input frozen images. However, using these methods, the generation of detailed texture and content inside the nuclei region, as it should appear in permanent section images, is still a challenging task, missing important diagnostic details. Additionally, these approaches generate content even in the blank regions of the frozen sections, introducing artificial information that could potentially mislead pathologists.

In this paper, we propose a new approach to enhance frozen section images by guidance of permanent section images using a combination of CycleGAN and a unique attention method. This new approach, called Segmented Attention Network (SAN), is based on nuclei segmentation that pushes the model to learn the details inside the nuclei region as they appear in permanent images, enhancing the frozen images to have richer data, especially in the nuclei area, yet without adding misleading details to the empty areas in the frozen section images. To obtain this, we developed a two-step training procedure, incorporating both the original pair of images and the nuclei-segmented pairs. While the task of the first step is to translate frozen images into permanent images, the second step acts as a hard-attention mechanism, transferring nuclei-segmented frozen images into nuclei-segmented permanent images. By incorporating nuclei-segmented pairs of frozen and permanent images in the training process and adding additional loss for this process, our approach is pushing the discriminator to focus more on the nuclei region, which, in turn, encourages the generator to generate more detailed content within this region, producing permanent images more accurately in these challenging regions.

# Methods

### Data collection and preprocessing

The data used to train our models contained breast, colon, and kidney cancer slide images, obtained from The Cancer Genome Atlas (TCGA), a publicly funded project initiated by the National Cancer Institute and the National Human Genome Research Institute. Overall, we used 46,912 pairs of frozen and permanent images for breast, 25,362 pairs for colon, and 13,691 pairs for kidney for training. These pairs were prepared for each cancer type, where each pair of frozen and permanent slides was acquired from the same patient to get the best correspondence as fully paired (same tissue slice), but exact frozen-permanent pairs were not available due to different preparation protocols. We used the images with the highest resolution and cropped each of them into patches of 256x256 pixels, while removing images with significant blank parts. Afterward, the data was shuffled.

### Segmented Attention Network (SAN) for frozen section image enhancement

Our goal is to utilize permanent images in order to improve frozen section images, placing particular emphasis on the nuclei regions to capture critical details with high accuracy.
The image-to-image translation from frozen images into permanent images was performed by CycleGAN.
We denote $f \in F$, where $f$ is a frozen section image taken from the frozen section set $F$, and $p \in P$, where $p$ is a permanent section image taken from the permanent section set $P$. The CycleGAN consists of two generators, $G_P$ and $G_F$, and two discriminators, $D_P$ and $D_F$. The generator $G_P$ learns to map $G_P : F \rightarrow P$, and the generator $G_F$ learns to map $G_F : P \rightarrow F$. Each of the generators has the architecture of Attention U-Net [26,27], as it improves the model sensitivity with minimal computation overhead. The two CycleGAN discriminators, $D_P$ and $D_F$, learn to distinguish between the real and generated images in their respective domains of permanent and frozen section images subset, respectively. Each of the discriminators has the architecture of Resnet-18 [28]. Our loss function is:

$$\mathcal{L}(G_F, G_P, D_F, D_P) = \mathcal{L}_{GAN}(G_P, D_P, F, P) + \mathcal{L}_{GAN}(G_F, D_F, P, F) + \lambda \mathcal{L}_{cyc}(G_P, G_F, F, P), \quad (1)$$

where $\mathcal{L}_{GAN}$ is the adversarial loss, $\mathcal{L}_{cyc}$ is the cycle-consistency loss, and $\lambda$ is a regulation weight to control the relative importance of the two objectives.

To improve the generation of frozen to permanent images specifically on the nuclei regions, we introduce a new loss function:

$$\mathcal{L}(G_F, G_P, D_F, D_P) = \mathcal{L}_{GAN}(G_P, D_P, F, P, F_{seg}, P_{seg}) + \mathcal{L}_{GAN}(G_F, D_F, P, F, P_{seg}, F_{seg}) + \lambda \mathcal{L}_{cyc}(G_P, G_F, F_{seg}, P_{seg}), \quad (2)$$

where $F_{seg}$ and $P_{seg}$ are the nuclei segmented images of the frozen and permanent images, respectively, which include only the nuclei contents in the image, as can be seen in Figure 2.

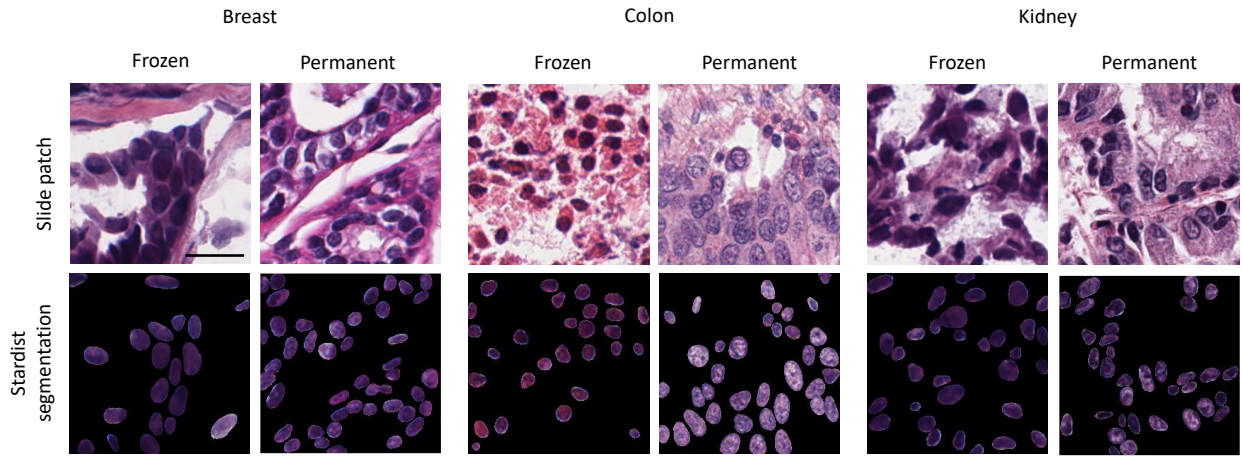

**Figure 2.** Examples of nuclei segmentation using StarDist on frozen and permanent patches. For the nuclei segmentation, we used the pre-trained StarDist model. Scale bar: 20 µm.

First, nuclei segmentation is done on the permanent and frozen patches using the StarDist model [29], pre-trained on H&E nuclei segmentation. Next, we train the model in two steps. The first step is done on the original pairs of $F$ and $G$, and the second step is done on the nuclei-segmented pairs of $F_{seg}$ and $P_{seg}$. See full algorithm flow on Table 1.

| **ALGORITHM FLOW** |
|---|
| *Inputs:*<br>*f, A batch of frozen images of 256×256 pixels,*<br>*p, A batch of permanent images of 256×256 pixels*<br><br>*Outputs:*<br>$G_F, G_P,$ *CycleGAN generators*<br>$D_F, D_P,$ *CycleGAN discriminators*<br><br>*For every epoch do:*<br>   *For every batch of {f, p} do:*<br>      *Generate segmented images:* $f_{seg}, p_{seg}$<br>      *Generate permanent images:* $\hat{p} \leftarrow G_p(f), \quad \hat{p}_{seg} \leftarrow G_p(f_{seg})$<br>      *Calculate the discriminator $D_P$ adversarial loss:* $\mathcal{L}_{D_P,GAN_{total}} = \mathcal{L}_{D_P,GAN} + \lambda_{seg} \cdot \mathcal{L}_{D_P,GAN_{seg}}$<br>      *Calculate the discriminator $D_F$ adversarial loss:* $\mathcal{L}_{D_F,GAN_{total}} = \mathcal{L}_{D_F,GAN} + \lambda_{seg} \cdot \mathcal{L}_{D_F,GAN_{seg}}$<br>      *Apply gradient descent on:* $\mathcal{L}_{D,GAN_{total}} = \mathcal{L}_{D_P,GAN_{total}} + \mathcal{L}_{D_F,GAN_{total}}$<br>      *Calculate the generator $G_P$ adversarial loss:* $\mathcal{L}_{G_P,GAN_{total}} = \mathcal{L}_{G_P,GAN} + \lambda_{seg} \cdot \mathcal{L}_{G_P,GAN_{seg}}$<br>      *Calculate the discriminator $G_F$ adversarial loss:* $\mathcal{L}_{G_F,GAN_{total}} = \mathcal{L}_{G_F,GAN} + \lambda_{seg} \cdot \mathcal{L}_{G_F,GAN_{seg}}$<br>      *Calculate cycle-consistency loss:* $\mathcal{L}_{cycle_{total}} = \mathcal{L}_{cycle} + \lambda_{seg} \cdot \mathcal{L}_{cycle_{seg}}$<br>      *Apply gradient descent on:* $\mathcal{L}_G = \mathcal{L}_{G_P,GAN_{total}} + \mathcal{L}_{G_F,GAN_{total}} + \lambda_{cycle} \cdot \mathcal{L}_{cycle_{total}}$ |

**Table 1.** Algorithm flow – Segmented Attention Network (SAN).

Applying the procedure appearing in Table 1 ensures that the discriminator will learn how to distinguish between the cell nuclei in the frozen-section domain and in the permanent-section domain in hard-attention manner, while pushing the generator to improve the cell nuclei details in the enhanced frozen section image. A block-diagram of the network flow is shown in Figure 3.

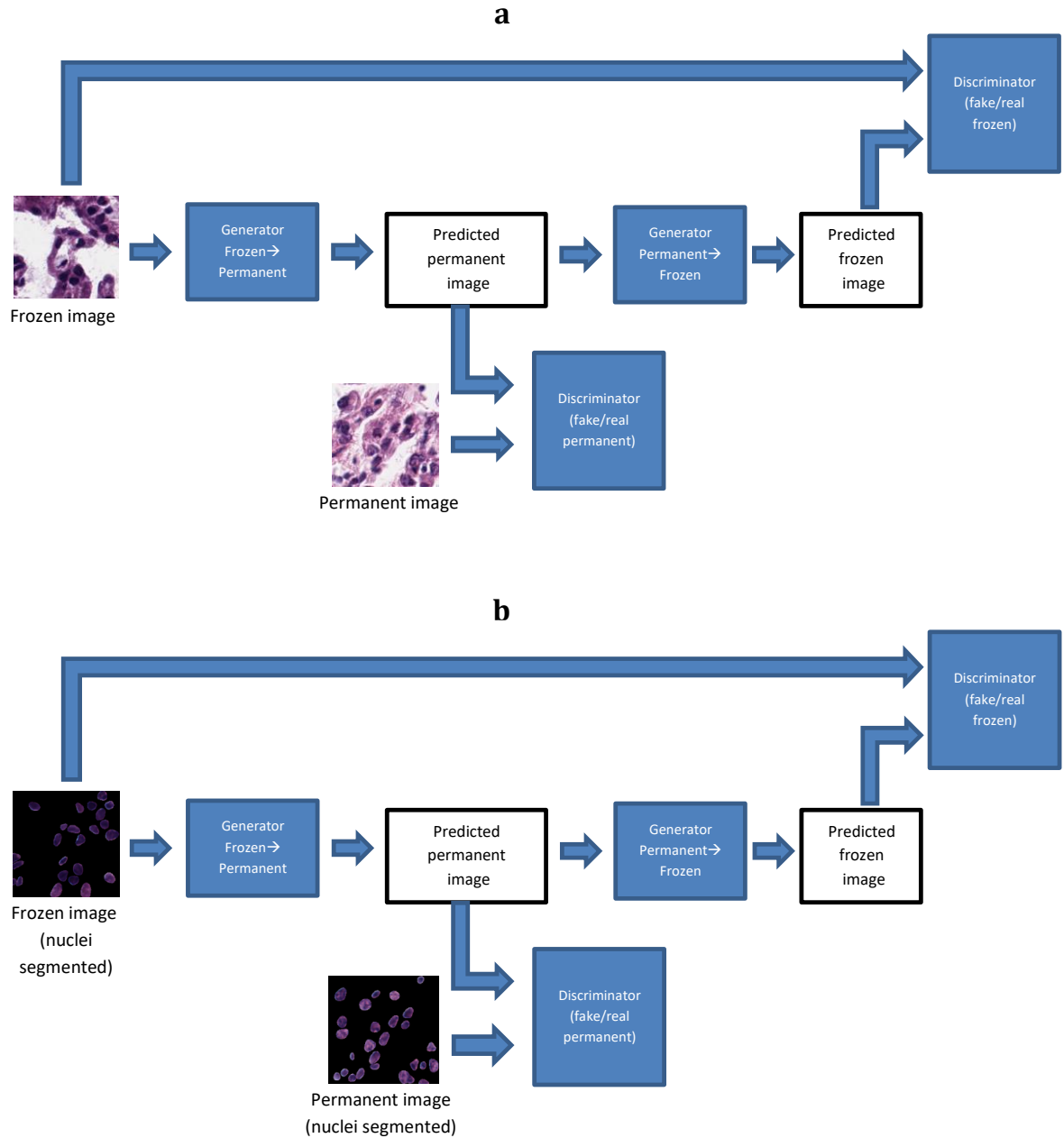

**Figure 3.** A block-diagram of the proposed method. (a) Generating a permanent-section-like image for a real frozen section image. The adversarial loss is set in a "minmax" configuration between the "fake" permanent section image and the un-paired "real" permanent section image, so at each step the generator and the discriminator improve. The correspondence of the generated permanent section image to the frozen section image input will be pushed using the cycle consistency loss. (b) The same generator is used to generate a nuclei-segmented permanent section image. The discriminator will then learn to distinguish between the "fake" nuclei-segmented permanent section image and the un-paired "real" nuclei-segmented permanent section image. At the same time, the generator learns to transform frozen-section nuclei to permanent-section nuclei, while pushing the generator attention to details in the nuclei region.

## Model configuration

We trained the models on ASUS TUF Dash F15 i7 PC, with NVIDIA GeForce RTX 3060 GPU. The model configuration including the hyper-parameters used for each method are presented in Table 2.

| Parameters | CycleGAN | CycleGAN with Attention UNET & Resnet | SAN (ours) |
|---|---|---|---|
| Image Input (pixels) | 256x256x3 | 256x256x3 | 256x256x3 |
| Batch Size | 1 | 1 | 1 |
| Epochs | 10 | 6 | 5 |
| Generator | Vanilla | Attention-UNET | Attention-UNET |
| Discriminator | Patch-GAN | Resnet-18 | Resnet-18 |
| Learning Rate | $1 \cdot 10^{-4}$ | Generator: $1 \cdot 10^{-4}$ Discriminator: $1 \cdot 10^{-5}$ | Generator: $1 \cdot 10^{-3}$ Discriminator: $1 \cdot 10^{-4}$ |
| $\lambda_{GAN}$ | 1 | 1 | 1 |
| $\lambda_{cycle}$ | 10 | 10 | 100 |
| $\lambda_{seg}$ | 0 | 0 | 1.0 |

**Table 2.** Model configuration for each method.

# Results

## Grad-cam evaluation

Our method pushes the discriminator to distinguish better on the feature of interest, which in our case is the nuclei region, and therefore to make the generator generate more reliable content in this region of interest. We first used Grad-cam [30] to examine the influence of our method on the discriminator. Grad-cams are a visualization method that highlights the regions of an image that contribute the most to a neural network's output, helping explain why a particular decision was made by the network. We compared CycleGAN with Attention UNET & Resnet, and SAN both with discriminator architecture of Resnet-18. We generated Grad-cams on the discriminator $D_P$. Figure 4 visually demonstrates that our method, SAN, puts more attention on the nuclei region, which pushes the generator to generate better nuclei content.

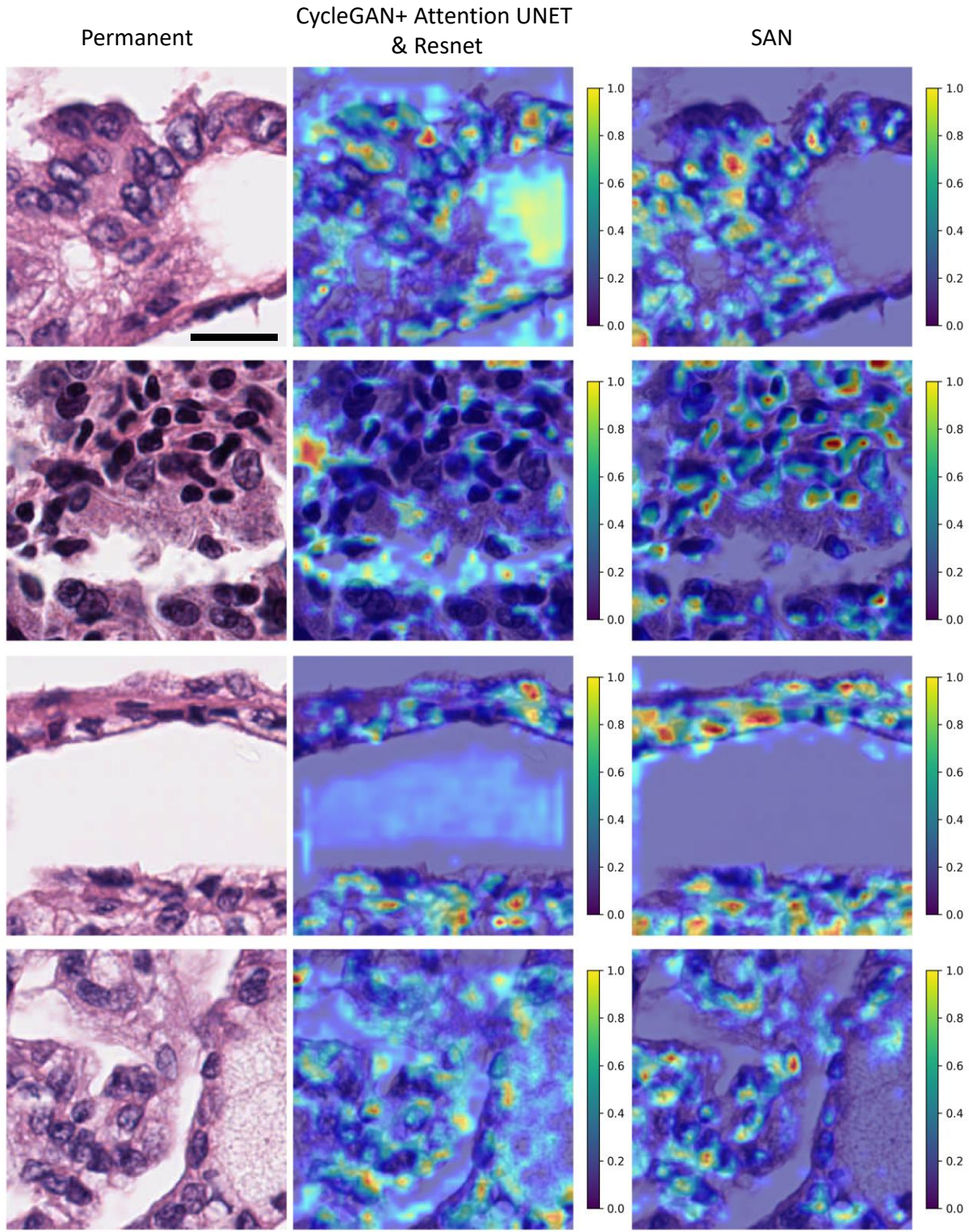

**Figure 4.** Grad-cam comparison. Examples of permanent section patches and the Grad-cam maps of the discriminator $D_P$. The Grad-cam maps on the second and third rows are for the discriminators of CycleGAN with Attention UNET & Resnet and SAN (our method), respectively. Using our method, SAN, the discriminator learns to put more focus on the nuclei parts of the image, while pushing the generator to put more efforts in generating more details in the nuclei regions as they should appear in a permanent section image. Scale bar: 20 μm.

## FID score evaluation

To evaluate the similarity of our generated permanent section images to the unpaired permanent section targets, we used Frechet Inception Distance (FID) [31], measuring the Wasserstein distance [32] between multivariate Gaussian distributions fitted to feature representations of the generated and real images, typically using features from an Inception v3 net trained on ImageNet. FID can be computed from the mean and covariance of Inception activations on the generated and real images. Lower FID indicates greater similarity between the generated and real image distributions, making it useful for evaluating improvements in image generation quality for GANs and other generative models. The FID score comparison between our method, the CycleGAN, and CycleGAN with Attention UNET & Resnet on the test data is presented in Table 3.

| Organ Tissue | Test data samples | CycleGAN | CycleGAN with Attention UNET & Resnet | SAN (ours) |
|---|---|---|---|---|
| Breast | 5,300 | 46.6 | 32.5 | **28.8** |
| Colon | 8,456 | 81.9 | 56.45 | **53.45** |
| Kidney | 7,436 | 79.32 | 63.85 | **51.7** |

**Table 3.** Comparison of FID scores between CycleGAN, CycleGAN with Attention UNET & Resnet and SAN (ours), using tissue sections from breast, colon, and kidney cancers.

## Visual inspection

First, to verify that our SAN model succeeds in generating both permanent section patches from frozen section patches and nuclei-segmented permanent section patches from nuclei-segmented frozen section patches using the same model, we visually inspected the full test data, with one example shown in Figure 5.

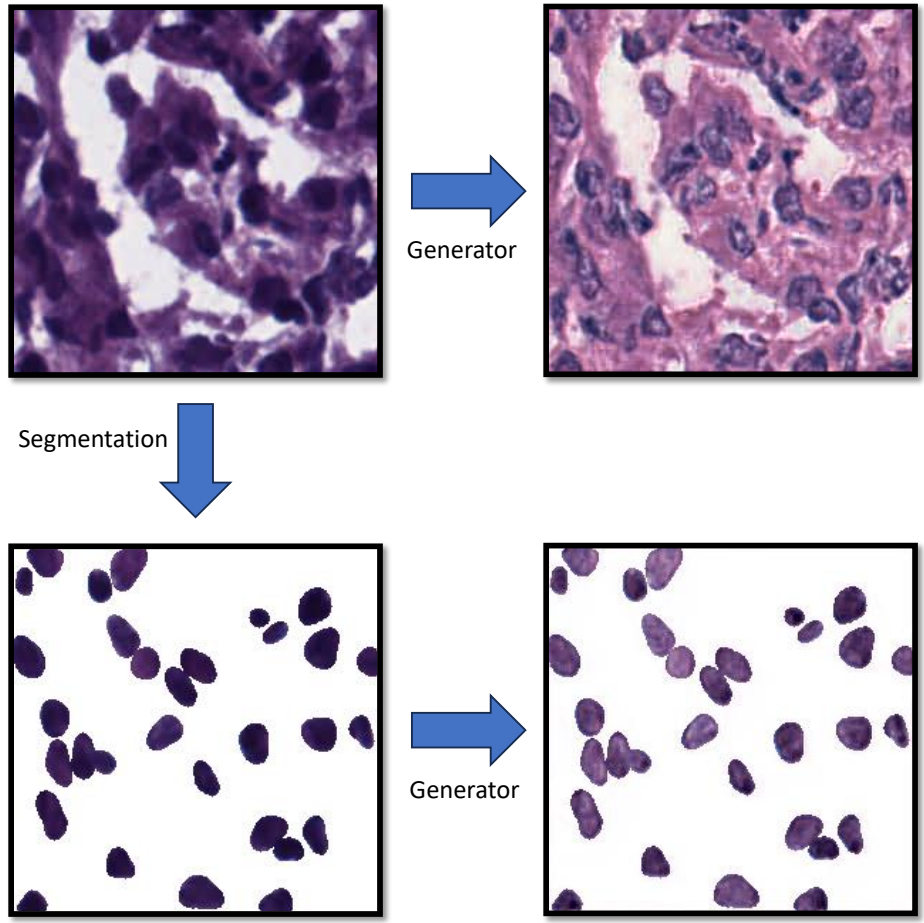

**Figure 5.** Example of generated permanent section patch from frozen section patch and nuclei-segmented permanent section patch from nuclei-segmented frozen section patch using the same model.

For a full comparison, we present visual comparisons of the generated permanent section images using the CycleGAN, CycleGAN with Attention UNET & Resnet, and SAN, focusing on the breast, colon, and kidney tissue samples. Examples of the generated permanent section images alongside their corresponding frozen section counterparts are provided in Figure 6. These examples illustrate the differences in image quality and detail preservation among the methods. CycleGAN tends to produce images with noticeable artifacts and a lack of fine details, particularly within the nuclei region. The CycleGAN with Attention UNET & Resnet shows a slight improvement in preserving details and avoiding artifacts, but still tends to inaccuracy in representing the structures within the tissue samples. In contrast, our proposed SAN method significantly enhances the frozen section, especially within the nuclei region, providing clearer and more diagnostically relevant images compared to the other methods.

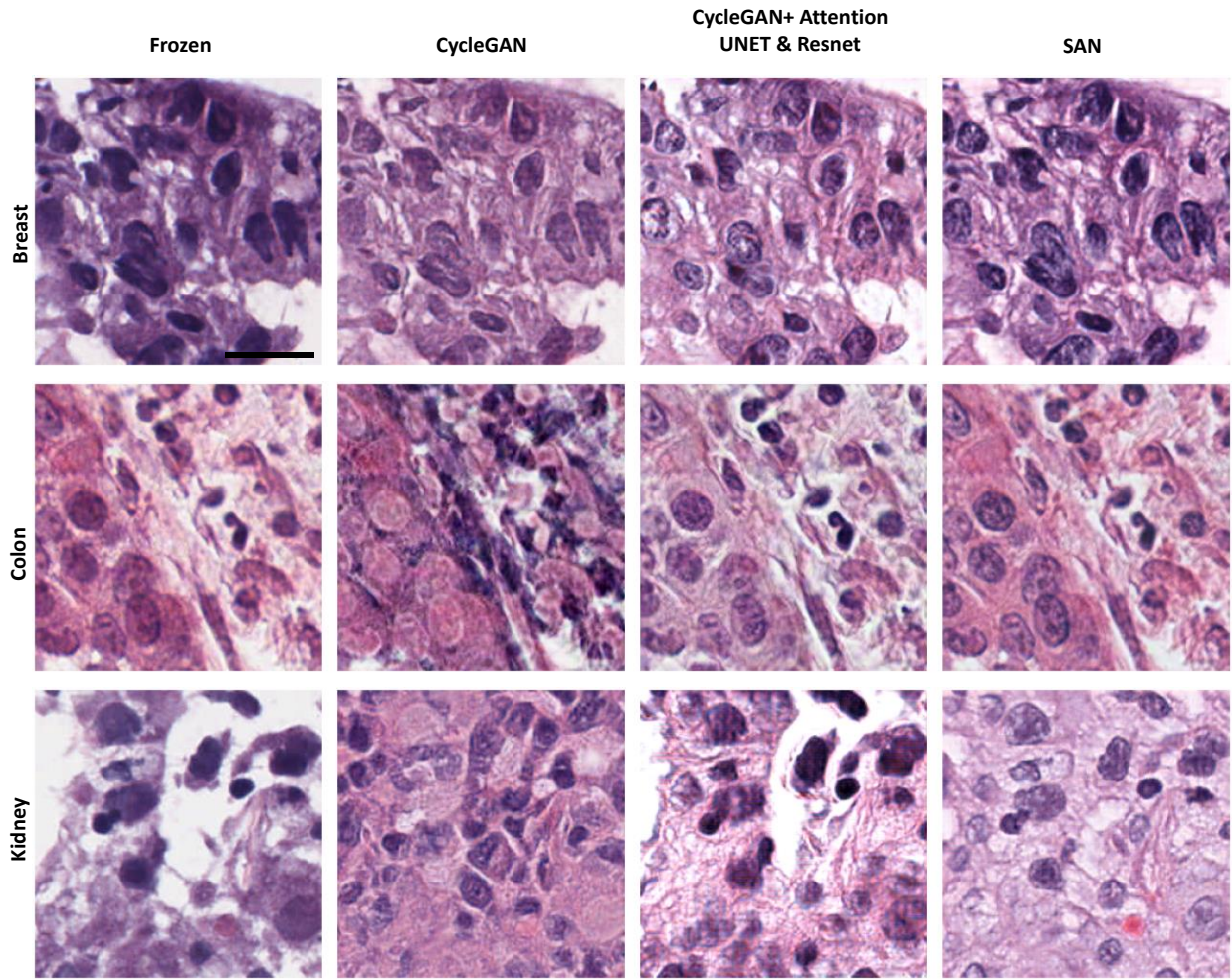

**Figure 6.** Example of generated breast permanent section images in zoom-in with CycleGAN, CycleGAN with Attention UNET & Resnet and with SAN (ours). Scale bar: 20μm.

## Nuclei texture analysis

To evaluate the content of the enhanced nuclei regions, we quantitatively analyzed the texture of kidney tissue images by comparing the segmentation of nuclei across frozen sections, permanent sections, and the enhanced frozen sections produced using our SAN method and CycleGAN with Attention UNET & Resnet. The focus is on extracting Gray-Level Co-occurrence Matrix (GLCM) [33] features, which are crucial for evaluating texture patterns related to diagnostic markers in histopathology.

For this analysis, we segmented the nuclei regions in all test images (frozen, permanent, and generated sections) and computed GLCM features from 14×14-pixel crops taken exclusively from within the nuclei. This ensured that our texture analysis focused on the most diagnostically relevant regions. We calculated GLCM features using distances of 1, 2, 3, 4, 5, and 6 pixels and angles of 0°, 45°, 90°, and 135°. The features analyzed include contrast, correlation, energy, and homogeneity, which provide a comprehensive statistical description of the textures within the nuclei.

The GLCM feature statistics are summarized in Table 4 below, which presents the mean and standard deviation (Std) for each feature across the permanent sections, frozen sections, CycleGAN with Attention UNET & Resnet generated images, and SAN enhanced images.

| Feature | Mean | | | | Std | | | |
| --- | --- | --- | --- | --- | --- | --- | --- | --- |
| | Permanent | Frozen | CycleGAN with Attention UNET & Resnet | SAN | Permanent | Frozen | CycleGAN with Attention UNET & Resnet | SAN |
| **Contrast** | 89.61 | 38.61 | 107.46 | 93.29 | 44.12 | 27.34 | 81.68 | 45.97 |
| **Correlation** | 0.86 | 0.89 | 0.86 | 0.85 | 0.07 | 0.08 | 0.09 | 0.08 |
| **Energy** | 0.09 | 0.15 | 0.09 | 0.09 | 0.04 | 0.07 | 0.02 | 0.02 |
| **Homogeneity** | 0.15 | 0.32 | 0.14 | 0.14 | 0.06 | 0.13 | 0.04 | 0.04 |

**Table 4.** GLCM features statistics of permanent sections, frozen sections, CycleGAN with Attention UNET & Resnet and SAN (ours) from kidney tissues.

These GLCM features provide important insights into the similarity of textures between the real permanent sections and the generated sections. Notably, the SAN method demonstrates superior performance, producing GLCM feature values that closely match those of the permanent sections, particularly in terms of contrast and energy, which are important for capturing diagnostic texture details. This can also be seen in the outline of the GLCM features histograms, presented in Figure 7.

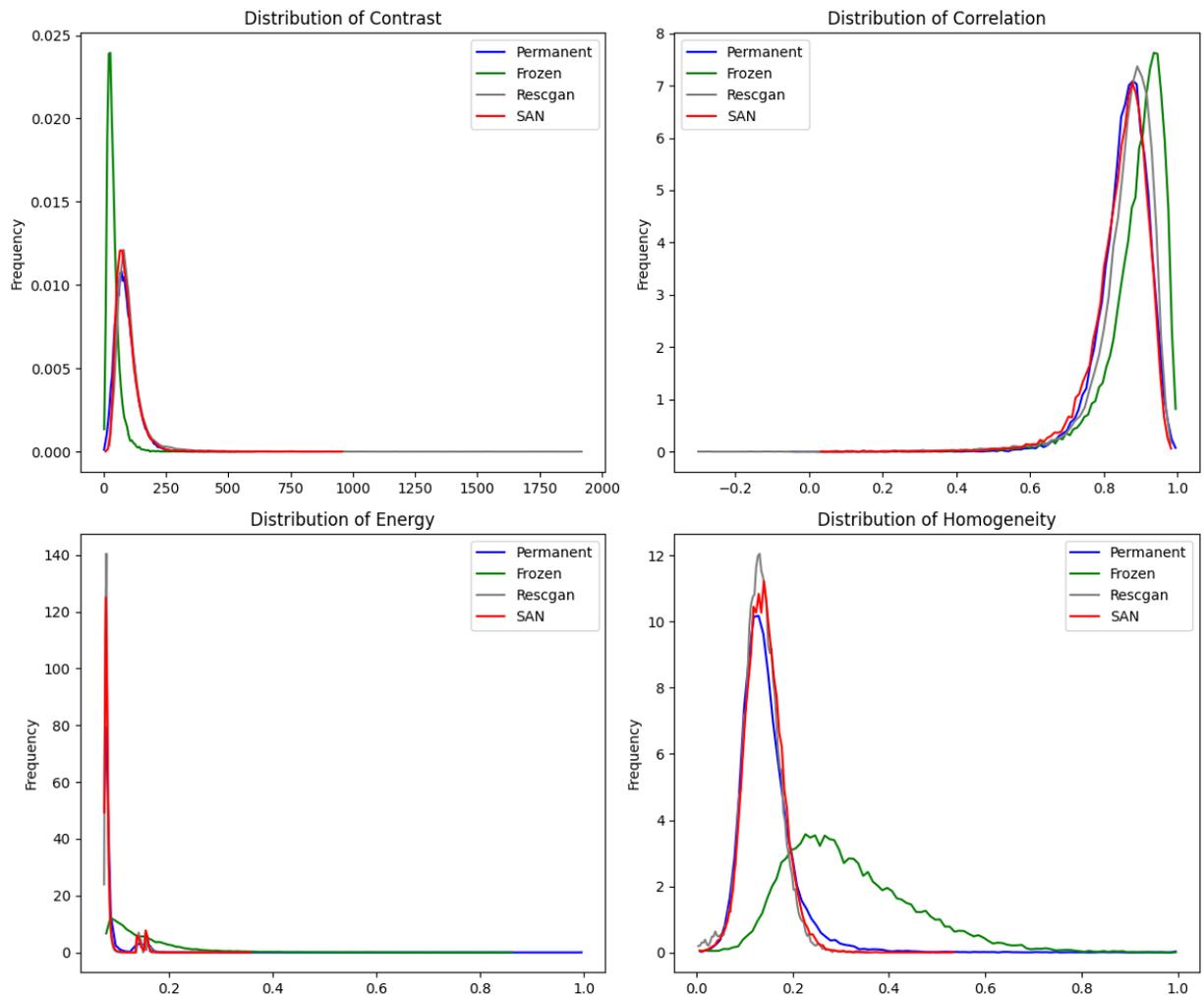

**Figure 7.** Histograms outline of the GLCM features: contrast, correlation, energy, and homogeneity of permanent sections, frozen sections, and the generated sections: CycleGAN with Attention UNET & Resnet and SAN (ours) from kidney tissues.

In addition to the feature statistics, we also computed the Jensen-Shannon (JS) divergence [34] to quantify the similarity of the GLCM feature distributions between each method and the real permanent sections. Table 5 presents the JS divergence for each GLCM feature:

| **Feature** | **Frozen** | **CycleGAN with Attention UNET & Resnet** | **SAN** |
|---|---|---|---|
| Contrast | 0.303645 | 0.416827 | 0.165063 |
| Correlation | 0.060312 | 0.077650 | 0.005735 |
| Energy | 0.350419 | 0.377750 | 0.277373 |
| Homogeneity | 0.340591 | 0.514835 | 0.325491 |
| **Average** | **0.263742** | **0.346765** | **0.193415** |

**Table 5.** JS divergence for GLCM features between permanent sections and frozen sections, CycleGAN with Attention UNET & Resnet and SAN (ours) from kidney tissues. SAN presents lower divergence from the real permeant sections.

The lower JS divergence values for SAN indicate that the textures generated by our method are closer to the real permanent sections. In particular, SAN achieves the lowest divergence in the image contrast feature (0.1651), indicating that it better reproduces the sharpness and variability of textures in the nuclei compared to both frozen sections and CycleGAN with Attention UNET & Resnet. In addition, SAN shows minimal divergence in the correlation feature (0.0057), suggesting that it accurately preserves the relationships between pixel intensities within the nuclei. Moreover, SAN exhibits improvements in the energy and homogeneity features over the other methods but still shows room for refinement in capturing smoothness and uniformity in the nuclei textures. Thus, the JS divergence analysis and GLCM feature statistics demonstrate that the proposed SAN method consistently outperforms both the frozen sections and CycleGAN with Attention UNET & Resnet in replicating the textures of the real permanent sections in the nuclei area. The significant reduction in divergence for contrast and correlation highlights SAN ability to capture fine textural variations within the nuclei, which are crucial for accurate tissue diagnosis. In contrast, CycleGAN with Attention UNET & Resnet shows higher divergence across all features, particularly for contrast and homogeneity, suggesting that it struggles to maintain the necessary texture details within the nuclei regions. Frozen sections also show relatively high divergence, especially in the contrast feature, further illustrating the limitations of the direct use of frozen section images for diagnostic purposes without enhancement.

To further assess the performance of our SAN model in a clinical setting, we generated 826 patches from frozen kidney tissue sections obtained at Chaim Sheba Medical Center. For each frozen section patch, we used our SAN model to produce an enhanced version. An example of such a pair is shown in Figure 8. We then asked a qualified pathologist (G. G.) to evaluate the contribution of the generated patches compared to the original frozen sections.

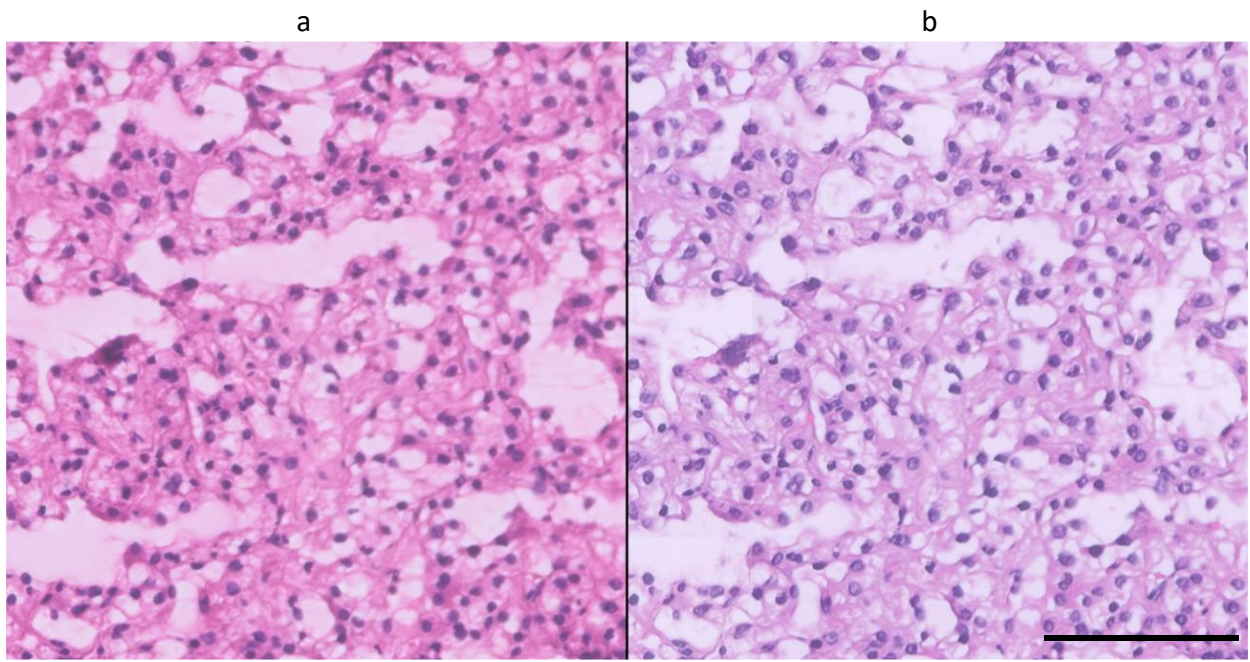

**Figure 8.** Example of frozen section kidney tissue patch (a) and its corresponding generated permanent section patch (b) produced by our SAN model, with visible texture enhancement inside the cell nuclei. Scale bar: 100 μm.

The most significant improvement is in the details that can be seen inside the nuclei. Both the boundaries of the nucleus and the dispersion of the chromatin and the presence of nucleoli are visible in the generated images. This feature is important for tissue grading, but also for the accuracy of diagnosis in a variety of different tumors. For example, when diagnosing papillary thyroid cancer, the pathologist might look for grooving inside the nucleus, which is sometimes difficult to see because of the quality of the image, and improving the quality of the nucleus can be critical [5, 6]. In our case, tissue grading of kidney cancer is based on the ability to see nucleation at different magnifications [5]. In relation to the cytoplasm and the borders between the cells, Figure 8 demonstrates a significant improvement. Regarding the cell borders, there is clearing of the cytoplasm and relatively clear cell boundaries in a way reminiscent of the permanent sections, instead of a smear of pink color, which characterizes the frozen sections. It appears that the model performs better to produce a permanent-section-like patches in areas where there are fewer artifacts of the frozen section. As the image has fewer tears in the tissue and more hypercellular areas where the concentration of cells is higher, it is more likely that the generated permanent-section-like patch will be of higher quality.

Quantitatively, according to the pathologist's analysis of the 826 kidney tissue patches: 74.9% of the generated section patches significantly improve the diagnosis-relevant nuclear details. 49.1% of the generated section patches produced clearing of the cytoplasm, and 59.7% produced clear cell borders as they should appear in the permanent section images.

## Discussion and Conclusion

In this paper, we introduced a new approach, called SAN, designed to address image-to-image translation with guided attention to specific details, and applied it to the challenging task of enhancing frozen section images by utilizing permanent section image guidance with focus on the cell nuclei details. While frozen sections offer immediate diagnosis during surgeries, but often suffer from artifacts and poor image quality, making it challenging to provide an accurate diagnosis, permanent sections offer better image quality and preservation of tissue architecture, but are time-consuming to prepare. Our approach aimed to bridge this gap by integrating a unique attention mechanism, based on nuclei segmentation, into the CycleGAN model. This frozen-section image enhancement is done by pushing the model generator to provide detailed content within the nuclei regions. Our method introduced a two-step training procedure, incorporating both the original pair of images and the nuclei-segmented pairs. While the first step translates frozen section images into permanent section images, the second step acts as a hard-attention mechanism, transferring nuclei-segmented frozen images into nuclei-segmented permanent images, which leads to more accurate and detailed images that capture nuclei-level information. Our method was evaluated using data samples from breast, kidney, and colon tissues. The unique attention method based on nuclei-segmentation significantly improved the generator's ability to capture and produce nuclei-level details, which are critical for precise diagnosis.

The proposed SAN method builds upon the framework of CycleGAN but incorporates attention mechanisms to prioritize the fidelity of nuclei features in the translated images. To further highlight the effectiveness of our approach, we conducted a comparative analysis of our method, SAN, with CycleGAN, and CycleGAN with attention UNET & Resnet. Our approach demonstrated its advantage in enhancing frozen section images with richer content in the nuclei region. Moreover, we used Grad-cam maps to visualize the impact of our method. The Grad-cam maps illustrated that our approach pushed the discriminator to focus more on the nuclei region, which, in turn, encouraged the generator to generate more detailed content within this region, producing permanent section images more accurately in these challenging regions. In addition, we evaluated the clinical improvements based on pathologist examination. While our results demonstrate improvements in image quality and detail preservation, several important considerations merit discussion. First, it is crucial to acknowledge the inherent challenge of working with unpaired frozen and permanent section images. We have obtained image pairs from the same patient, but the frozen and permanent section images are not exactly paired due to different chemical preparation protocols. The lack of direct correspondence between the two types of sections poses a fundamental limitation. As a result, there is ambiguity regarding whether the model has learned features from cancerous or non-cancerous areas, as the presence or absence of cancerous tissue areas may vary within individual patients. Addressing this issue would require meticulous labeling by pathologists to confirm the characteristics of each image pair, a task that we intend to explore in future work. Second, ensuring the relevance of enhancement learned from permanent section patches to corresponding frozen section patches is essential for the effectiveness of our translation model. While we have attempted to mitigate this concern through careful data selection and training, it remains a challenge to guarantee the alignment of features between the two types of sections. To address this, incorporating pathologist labeling can be carried out to validate the suitability of each image pair for training, thereby enhancing the relevance and accuracy of our model learning process.

Third, the discrepancy in morphological characteristics between frozen and permanent sections presents a significant obstacle in achieving fully paired data. While it is not feasible to eliminate this difference entirely, we recognize the importance of utilizing successive or closely positioned histological section paired data to refine our model further. By exploring different degrees of pairing and conducting additional model training on such data, it will be possible to optimize the performance of our approach and enhance its robustness in translating frozen images into permanent-like quality.

In conclusion, SAN represents a significant advancement in the field of histopathology image enhancement. By combining nuclei segmentation with CycleGAN, we have developed a general method, which specifically addresses the limitation of frozen section images and offers a fast and reliable solution for frozen section image enhancement. This approach has the potential to improve the accuracy and effectiveness of pathological diagnosis and can be extended to various applications where detailed attention to specific image regions is required.

# Supporting information

Not applicable.

# Conflict of interest

# Code availability

The code is available upon request.

# Data availability statement

The slide images we used to train and test our models were taken from the TCGA Research Network: https://www.cancer.gov/tcga
In the Clinical Analysis section, we used for evaluation 826 patches taken from 4 kidney tissues obtained at Chaim Sheba Medical Center.

# Funding

Supported by the Israel Science Foundation (ISF).

# Keywords